\documentclass[final,1p,times]{elsarticle}

\usepackage{amssymb}

 \usepackage{pgfplots}
 \usepackage{pgfplotstable}
 \usepackage{tikz}
 \usepackage{tkz-euclide} 
 \usetkzobj{all} 
 \usepackage{tkz-berge}
\usepackage{./subfigure}
\usepackage[colorlinks=true]{hyperref}
\usepackage{./algorithm,./algorithmic}
\newfloat{algorithm}{thp}{loa}[section]

\journal{Journal of Computational Physics}

\def\dt{{\delta t}}

\begin{document}

\begin{frontmatter}



\title{A ``Necklace'' Model for Vesicles Simulations in 2D}


\author[label1]{Mourad Ismail\corref{cor1}}
\ead{Mourad.Ismail@ujf-grenoble.fr}
\author[label2]{Aline Lefebvre-Lepot}
\ead{Aline.Lefebvre@polytechnique.edu}
\address[label1]{Universit\'e Grenoble 1 / CNRS, Laboratoire Interdisciplinaire de Physique / UMR 5588, Grenoble, F-38041, France}
\address[label2]{CNRS/CMAP - Ecole Polytechnique - Route de Saclay - 91128 Palaiseau Cedex, France}
\cortext[cor1]{Corresponding author}

\begin{abstract}

  The aim of this paper is to propose a new numerical model to
  simulate 2D vesicles interacting with a newtonian fluid. The
  inextensible membrane is modeled by a chain of circular rigid
  particles which are maintained in cohesion by using two different
  type of forces. First, a spring force is imposed between neighboring
  particles in the chain. Second, in order to model the bending of the membrane,
  each triplet of successive particles is submitted to an angular
  force.

  Numerical simulations of vesicles in shear flow have been run using
  Finite Element Method and the
  \texttt{FreeFem++}\cite{hecht2010finite} software. Exploring
  different ratios of inner and outer viscosities, we recover the well
  known ``Tank-Treading'' and ``Tumbling'' motions predicted by theory
  and experiments. Moreover, for the first time, 2D simulations of the
  ``Vacillating-Breathing'' regime predicted by theory in
  \cite{PhysRevLett.96.028104} and observed experimentally in
  \cite{PhysRevLett.98.128103} are done without special ingredient
  like for example thermal fluctuations used in \cite{PhysRevE.80.011901}.
\end{abstract}


\end{frontmatter}


\section{Introduction}
\label{sec:intro}

A vesicle is a closed phospholipid membrane, separating an internal
fluid from the external suspending medium. Studying collections of
vesicles embedded in a flow is of great interest, since such systems
consist in an efficient and flexible model for much complex flows as
blood  (see \cite{pozrikidis1990axisymmetric,noguchi2005shape,kaoui2009red}
for example). Of course, the simplicity of the model cannot reproduce
the complexity of the biological and biochemical activities of living
cells. However, vesicles can be considered as a simple model to study
some mechanical properties of Red Blood Cells, which has a great
influence on their global comportment.

From the numerical point of view, vesicle simulations have been
carried out using several numerical methods. We can mention the
molecular dynamics models \cite{markvoort2006vesicle}, the boundary
integral methods in unbounded domains
\cite{beaucourt2004steady,kaoui2008lateral,ghigliotti2010rheology} and
recently in confined domains
\cite{veerapaneni2009boundary,rahimian2010dynamic,ghigliotti2011vesicle}.  There is also
numerical studies of vesicle dynamics based on finite difference
methods coupled with level set techniques (see
\cite{maitre2009applications,salac2011level} for example).

To the best our knowledge, finite element methods are seldom used to
simulate vesicle dynamics. However, we can cite \cite{bui2009dynamics}
for a contact algorithm coupled with finite element and these recent
works
\cite{LAADHARI:2011:HAL-00604145:2,DOYEUX:2011:HAL-00665007:1,DOYEUX:2012:HAL-00665481:1}
for coupling Level Set with Finite Element.

Apart from level set techniques, we can also mention the Phase Field method in the class of capture interface methods used to simulate vesicles.
See for example \cite{wang2008modelling,PhysRevE.79.031926} or
\cite{du2004analysis} for a numerical analysis of this method.

We can also refer to the numerical method developed in
\cite{Tsubota2010} for Red Blood Cells simulations and where the main
ingredient is to introduce two energies for elasticity and bending in order
to model the RBC membrane. The whole problem, made of fluid and
membrane interaction, is solved using a moving particle semi-implicit
method. In the same context, we find in \cite{PhysRevE.80.011901} a
theoretical and numerical study of 2D vesicles under shear. The
numerical method is based on a mesoscale simulation technique in which
the membrane is modeled as a set of monomers connected thanks to a
spring potential. The bending of the membrane is also taken into
account using a bending potential. Theses potentials are similar to
the energies used in \cite{Tsubota2010}. The latter work is
particularly interesting because it covers for the first time the
Vacillating-Breathing motion in 2D. This is done by adding a thermal
fluctuation to the membrane model.

In this paper, we study the dynamic of a 2D vesicle in shear flow by
introducing a new model. This method consists in modelling the
vesicle's membrane by a ``necklace'' of circular rigid particles which
are maintained in cohesion by using a spring force imposed between
neighboring particles in the chain. Regarding the bending of the
membrane, it is modeled by imposing an angular force on each triplet
of successive particles. These two forces are derived form energies
which are somewhat similar to those used in \cite{PhysRevE.80.011901}
and \cite{Tsubota2010}.

The whole problem is solved using Finite Element Method and the
\texttt{FreeFem++}\cite{hecht2010finite} software. From a numerical
point of view, the rigidity of each particle of the the chain is
imposed using a penalty method, which physically consists in making
the viscosity go to infinity in the rigid domain
(see~\cite{janela2005penalty,lefebvre2007fluid}). In order to avoid
particles overlapping, we generalize the inelastic contact model
proposed in~\cite{lefebvre2007fluid}. Finally, a constraint is added,
in order to impose the vesicle to have a constant area.

This paper is organized as follow : First, we introduce our model for
the vesicle's membrane and its coupling with Stokes equations. Second,
we describe the numerical method and the way to impose the different
constraints (rigid-body motion, constraint of non overlapping and
constant area of the vesicle). Third, we introduce some relevant
physical parameters and we finish by some numerical results to
validate our model. This validation is done by using the well known
behaviors of vesicles : equilibrium shapes, ``Tank-Treading'' (TT)
and ``Tumbling'' (TB) motions. Finally, we present the transition
between TT and TB regimes and we recover the ``Vacillating-Breathing''
(VB) (also called ``Swinging'' (SW)) motion in 2D.

\section{Modeling}

Let's recall that a vesicle is a system of two fluids separated by a
membrane. In this work, we are in two dimensions and we denote by
$\Omega$ a bounded domain in ${\mathbb R}^2$ that represents the whole
system (internal and external fluids and the membrane).

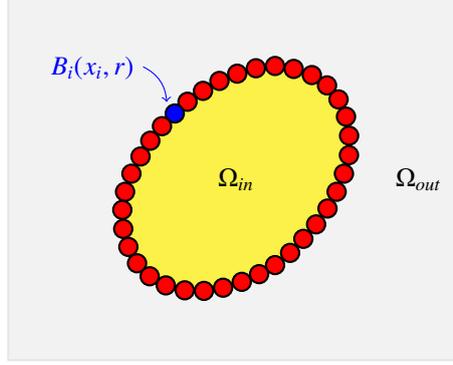
\begin{figure}[!h]
  \centering
  \begin{tikzpicture}[scale=0.6]
  \def\Radius{0.2 cm}
  \draw[black,thick,fill=gray, opacity=0.1] (-5, -4) rectangle (5, 4);  
  
  \draw[rotate=45,fill=yellow ,opacity=0.7] (0,0) ellipse (2.9cm and 2cm);
  
  \node[black] at (0:0cm){$\Omega_{in}$};
  \node[black] at (0:4cm){$\Omega_{out}$};

  \node (a) [blue] at (5cm:4cm){$B_i({x}_i,r)$};
  \node (b) at (-1.4, 1.5) {};
  \path[->] [blue] (a) edge [bend left]  (b);
    
  \draw[black,thick,fill=red] ( 1.47 ,-1.32) circle (\Radius);
  \draw[black,thick,fill=red] ( 1.75 ,-1) circle (\Radius);
  \draw[black,thick,fill=red] ( 2    ,-0.65) circle (\Radius);
  \draw[black,thick,fill=red] ( 2.2  ,-0.28) circle (\Radius);
  \draw[black,thick,fill=red] ( 2.36 , 0.12) circle (\Radius);
  \draw[black,thick,fill=red] ( 2.47 , 0.53) circle (\Radius);
  \draw[black,thick,fill=red] ( 2.47313, 0.96) circle (\Radius);
  \draw[black,thick,fill=red] ( 2.40542, 1.380521) circle (\Radius);
  \draw[black,thick,fill=red] ( 2.24182, 1.761079) circle (\Radius);
  \draw[black,thick,fill=red] ( 1.98884, 2.071429) circle (\Radius);
  \draw[black,thick,fill=red] ( 1.65657, 2.299197) circle (\Radius);
  \draw[black,thick,fill=red] ( 1.25826, 2.435304) circle (\Radius);
  \draw[black,thick,fill=red] ( 0.85, 2.5) circle (\Radius);
  \draw[black,thick,fill=red] ( 0.43, 2.45) circle (\Radius);
  \draw[black,thick,fill=red] (  0.02, 2.33) circle (\Radius);
  \draw[black,thick,fill=red] ( -0.37, 2.17) circle (\Radius);
  \draw[black,thick,fill=red] (-0.73, 1.95) circle (\Radius);
  \draw[black,thick,fill=red] (-1.07, 1.71) circle (\Radius);
  \draw[black,thick,fill=blue] (-1.35, 1.45) circle (\Radius);
  \draw[black,thick,fill=red] (-1.63, 1.17) circle (\Radius);
  \draw[black,thick,fill=red] (-1.88626, 0.848255) circle (\Radius);
  \draw[black,thick,fill=red] (-2.1, 0.5) circle (\Radius);
  \draw[black,thick,fill=red] (-2.3, 0.12) circle (\Radius);
  \draw[black,thick,fill=red] (-2.44,-0.28) circle (\Radius);
  \draw[black,thick,fill=red] (-2.5,-0.68) circle (\Radius);
  \draw[black,thick,fill=red] (-2.48,-1.1) circle (\Radius);
  \draw[black,thick,fill=red] (-2.36734,-1.498070) circle (\Radius);
  \draw[black,thick,fill=red] (-2.17679,-1.859579) circle (\Radius);
  \draw[black,thick,fill=red] (-1.89947,-2.146953) circle (\Radius);
  \draw[black,thick,fill=red] (-1.54642,-2.348735) circle (\Radius);
  \draw[black,thick,fill=red] (-1.13171,-2.456879) circle (\Radius);
  \draw[black,thick,fill=red] (-0.71,-2.47) circle (\Radius);
  \draw[black,thick,fill=red] (-0.29,-2.4) circle (\Radius);
  \draw[black,thick,fill=red] ( 0.12,-2.27) circle (\Radius);
  \draw[black,thick,fill=red] ( 0.5,-2.1) circle (\Radius);
  \draw[black,thick,fill=red] ( 0.85,-1.9) circle (\Radius);
  \draw[black,thick,fill=red] ( 1.18,-1.63) circle (\Radius);
\end{tikzpicture}
  \caption{Necklace model of the vesicle.}
  \label{fig:notations}
\end{figure}

We model the vesicle membrane by a
``necklace'' of $N$ rigid circular particles (See
figure~\ref{fig:notations}). We denote by $(B_i)_{i=1\ldots N}$ these
particles, $x_i$ their centers and $r$ their common radius. The domain
$\Omega$ is devided into three sub-domains: $B=\cup_{i=1}^N B_i$ is the
whole rigid domain (rigid by parts), and $\Omega_{in}$ and
$\Omega_{out}$ correspond to the internal fluid and the external
suspending medium respectively. These two last domains are supposed to
be filled with fluids governed by Stokes equations with respective 
viscosities $\mu_{in}$ and $\mu_{out}$.

In order to model properly the membrane, let's recall the main
properties of vesicles
\begin{itemize}
\item \textbf{Bending Energy : } it costs some energy to bend the
  membrane. This energy has to be taken into account in the model
\item \textbf{Local Inextensibility :} the membrane has to keep the
  same area (perimeter in 2D) during the simulation time
\item \textbf{Constant Volume :} the vesicle membrane is often
  considered to be impermeable. Moreover, inner and outer fluids are
  incompressibles so the vesicle volume (area in 2D) must stay constant.
\end{itemize}

\begin{figure}[!h]
  \centering
  \hspace*{3cm}
\begin{tikzpicture}[scale=0.45,rotate=0.25]
  \def\Radiuse{2.1 cm}
  \tikzstyle{spring}=[thick,decorate,decoration={coil,
    pre length=0.35cm,post length=0.35cm,aspect=0.6,amplitude=7}]

  \tkzDefPoint[label=below:$O$](0,0){o}
  \tkzDefPoint[label=above:$B_{i-1}$](8,4.12){pim}
  \tkzDefPoint[label=right:$B_{i}$](9,0){pi}
  \tkzDefPoint[label=below:$B_{i+1}$](8,-4.12){pip}
  \tkzDefPoint(10.2,4.5){pp}
  
  \draw[gray,thick,<->] ($(pi)+(1,-.4)$) to ($(pip)+(1,-0.4)$);
  \node[right] at (9.5,-2.46) {$\ell_i$};

  \draw[black,thick,->] ($(pi)+(3,-1)$) to ($(pip)+(3,-1)$);
  \node[right] at (11.5,-3) {$e_{i+1}$};
  \draw[black,thick,->] ($(pim)+(3,0.4)$) to ($(pi)+(3,0.4)$);
  \node[right] at (11.5,3) {$e_i$};
  
  \Edge[label=$R$](o)(pim)
  \Edge[label=$R$](o)(pi);
  \Edge[label=$R$](o)(pip);
  
  \draw[black,thick] (pim) circle (\Radiuse);
  \draw[black,thick] (pi) circle (\Radiuse);
  \draw[black,thick] (pip) circle (\Radiuse);

  \draw[dashed,line width=1pt] ($(pip)!4cm!(pi)$) -- ($(pip)!8.5cm!(pi)$);
  
  \draw[spring,black] (pip) -- (pi);
  \draw[spring,black] (pi) -- (pim);

  \tkzMarkAngle(pip,o,pi)
  \node[left,blue] at ($(o)+(2.3,-.7)$) {$\theta_i$};
  \tkzMarkAngle(pi,o,pim)
  \node[left,blue] at ($(o)+(2.3,.25)$) {$\theta_i$};
  \tkzMarkAngle[fill=blue!70](pp,pi,pim)
  \draw[blue,<-] ($(pi)+(0.1,1.1)$) to[out=110,in=-130] ($(pp)-(0.5,2)$) node[right] {$\theta_i$};

  \tkzMarkAngle[size=6mm](pi,pip,o)
  \node[left,red] at ($(pip)+(-.3,.6)$) {$\beta_i$};
  \tkzMarkAngle[size=6mm](o,pi,pip)
  \node[left,red] at ($(pi)-(.5,.5)$) {$\beta_i$};
  \tkzMarkAngle[size=6mm](pim,pi,o)
  \node[left,red] at ($(pi)+(-.5,.3)$) {$\beta_i$};
  \tkzMarkAngle[size=6mm](o,pim,pi)
  \node[left,red] at ($(pim)+(-0.2,-0.9)$) {$\beta_i$};
\end{tikzpicture}
  \caption{Notations.}
  \label{fig:notations_capillaire}
\end{figure}
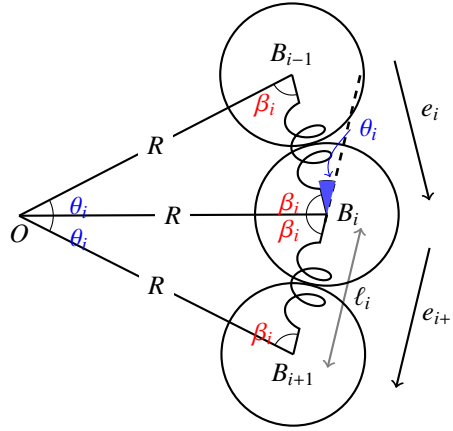

To satisfy the first two points, we introduce a couple of energies :
\begin{description}
\item[A bending energy] which consists in submitting each triplet of rigid
  (successive) particles to an angular force that tends to align them
  (see figure \ref{fig:notations_capillaire}). Therefore, ideally
  (when the vesicle is not deflated) the shape of the membrane is
  circular (spherical in 3D). This energy may be written as
  \begin{equation}
    \label{eq:1}
    E_b=\sum_{i}k_{a_i} (e_i \cdot e_{i+1} + 1),
  \end{equation}
  where $k_{a_i}$ is the spring constant acting on
  $(B_{i-1},B_i,B_{i+1})$ triplet, $e_i$ is the unit vector
  between particles $B_{i-1}$ and $B_i$ and $e_{i+1}$ is the unit vector
  between particles $B_i$ and $B_{i+1}$. These vectors are given by
  (see figure \ref{fig:notations_capillaire}) 
  \begin{equation}
    \label{eq:2}
    e_i = \frac{x_i - x_{i-1}}{|x_i - x_{i-1}|}, \quad
    e_{i+1} = \frac{x_{i+1} - x_i}{|x_{i+1} - x_i|}.
  \end{equation}
  In the following we assume that all -- angular -- springs have the
  same constant $k_a$.
\item[A stretching energy] whose role is to keep sticking each pair of
  successive particles through an elastic force. We can simply write
  this energy in this form
  \begin{equation}
    \label{eq:3}
    E_{st}=\sum_i k_{{rp}_i} (\ell_i-\ell_{0_i})^2,
  \end{equation}
  where $k_{{rp}_i}$ stands for the spring constant between particles
  $(B_i,B_{i+1})$ (cf. figure \ref{fig:notations_capillaire}), $\ell_{0_i}$ for its free-length and
  $\ell_i$ for the length of the $i^{th}$ spring. As in
  the case of bending energy, we assume that all springs have the same
  constant $k_{{rp}_i}=k_{rp}$ and free-length $\ell_{0_i}=\ell_0$.
\item[The Constant Area] is insured by imposing a constraint which will be
  described in section \ref{sec:extra}.
\end{description}

The forces derived from these energies will be plugged in the right
hand side of fluids model. This gives the major advantage to decouple
entirely fluid solver and membrane model. Thus, the
unknowns of the problem are the velocity and pressure fields in
$\Omega\setminus\bar B$, respectively denoted by $u=(u_1,u_2)$ and
$p$, together with the translational and angular velocities of the
rigid particles, denoted by $V=(V_1,\ldots,V_N)\in\mathbb{R}^{2N}$ and
$\omega=(\omega_1,\ldots,\omega_N)\in\mathbb{R}^{N}$. 

Most of known behaviors of vesicles under flow take place at very low
Reynolds number so it is natural to assume that, at each time $t$, the
fluid flow in $\Omega\setminus\bar B=\Omega\setminus\bar B(t)$ is
governed by Stokes equations with given Dirichlet boundary conditions
$u_0$ on $\partial \Omega$.
 \begin{equation}
\label{eq:Stokes}
\left\{
\begin{array}{rcll}
\displaystyle -\mu \triangle u+ \nabla p&=&0& \hbox{ in
}\Omega\setminus\bar B,\vspace{6pt}\\
\displaystyle \nabla \cdot u&=&0 &\hbox{ in } \Omega\setminus\bar B,\vspace{6pt} \\
\displaystyle u&=&u_0 &\hbox{ on } \partial\Omega
\end{array}
\right.
\end{equation}
where $\mu$ is equal to $\mu_{in}$ in $\Omega_{in}$ and $\mu_{out}$ in $\Omega_{out}$. 
In addition, we impose a no-slip boundary condition on $\partial B$:
\begin{equation}
\label{eq:noslipBC}
u(t,x) = V_i(t) + \omega_i(t) (x - x_i(t))^{\perp} \hbox{ on  } \partial B_i(t),\,\,\,
\forall i,
\end{equation}
where $x^{\perp}=(x_1,x_2)^{\perp}$ is the vector $(-x_2, x_1)$.
Finally, 
these equations are coupled
by writing the equilibrium of forces and torques on each $B_i$:
\begin{equation}
\label{eq:forces}
\left\{
\begin{array}{rcll}
\int_{B_i} f_i - \int_{\partial B_i}\sigma n & = & 0, & \hbox{  } \forall i,\vspace{6pt}\\
\int_{B_i}(x-x_i)^{\perp} \cdot f_i - \int_{\partial B_i}(x-x_i)^{\perp}\cdot \sigma n & = &0,  &\forall i,
\end{array}
\right.
\end{equation}
where $f_i$ is the volume force exerted on $B_i$
($f_i=(f^{st}_i+f^b_i)/|B_i|$), $n$ the outer normal to
$\Omega\setminus\bar B$ and $f^{st}_i$ end $f^b_i$ are the forces
derived from stretching and bending energies respectively. Finally,
recall that $\sigma$ stands for the stress tensor which can be written
as $\sigma=2\mu \hbox{D}(u) - p\hbox{I}$ where we denoted by
$\hbox{D}(u)$ the deformation tensor ($\hbox{D}(u) = \frac{\nabla u +
  (\nabla u)^t}{2}$).

A first difficulty to solve this problem is the dependence in time of each part of the domain ($\Omega_{in}$,
$\Omega_{out}$ and $B_i$) which imposes a priori to use time dependent conform meshes. A second one is the explicit coupling between the membrane model and the hydrodynamic forces. We will see in
section \ref{sec:num-methds} that we can deal with these difficulties using a Penalty method.

\section{Numerical methods}
\label{sec:num-methds}
\subsection{A first fluid/particle algorithm}

Instead of solving problem
(\ref{eq:Stokes})-(\ref{eq:noslipBC})-(\ref{eq:forces}) in it's
original form, we follow the method proposed in
\cite{janela2005penalty,lefebvre2007fluid}. More precisely, we use a
penalty method to impose the rigid motion on $B$. This allows us to
rewrite the problem initialy defined on a moving domain
$\Omega\setminus\bar B(t)$ into a global problem defined on $\Omega$.
Thus, the mesh of $\Omega$ that we consider can be fixed once for all
and does not change in time. Moreover, this allows us to handle the
fluids/rigid particles interaction implicitly rather than discretizing
problem~(\ref{eq:forces}). 
To do so, following a Fictitious Domain approach, 
we start by introducing the space $K_B$, made of functions defined on the whole domain $\Omega$, 
and in which we will search for a velocity field 
\begin{equation}
  \label{eq:4}
  K_B  =  \left\{ 
    \displaystyle v\in H^1_0(\Omega),\forall i\quad\exists
    (V_i,\omega_i) \in \mathbb{R}^2\times\mathbb{R},
    v = V_i + \omega_i(x-x_i)^{\perp} \hbox{ a.e. in
    }B_i
  \right\}.
\end{equation}
This space can be rewritten as follow (see \cite{janela2005penalty})
\begin{equation}
  \label{eq:5}
  K_B = \left\{
    v\in H^1_0(\Omega),\quad\hbox{D}(v)=0 \hbox{ a.e. in }B
  \right\}.
\end{equation}
We see clearly in this latter expression that elements of $K_B$
represent velocity fields in $\Omega$ which are subject to rigid body
constraint in $B$.  Moreover, it can be shown that
problem (\ref{eq:Stokes})-(\ref{eq:noslipBC})-(\ref{eq:forces}) is
equivalent to the following variational formulation in the whole
domain $\Omega$ (see \cite{janela2005penalty} for more details) :
\begin{equation}
({\cal P})\quad\left\{
\begin{array}{l}
\hbox{Find } u \hbox{ in } K_B \hbox{ and } p  \hbox{ in }  L^2_0(\Omega) \hbox{ such that }\vspace{6pt}\\
\displaystyle
2\mu \int_\Omega \hbox{D}(u) : \hbox{D}(\tilde u) 
- \int_\Omega p \nabla\cdot\tilde u
= \int_\Omega f \cdot \tilde u
,\,\,\, \forall 
 \tilde u \in K_B,\vspace{6pt}\\
\displaystyle\int_\Omega q\nabla \cdot u = 0 ,\,\,\,
\forall q\in L^2_0(\Omega)  , \\
\end{array}
\right. 
\end{equation}
where $L^2_0(\Omega)$ stands for the set of functions in $L^2(\Omega)$
with zero-mean and  $\displaystyle f = \sum_{i=1}^N f_i
\mathbf{1}_{B_i}$ ($\mathbf{1}_{B_i}$ denotes the characteristic
function of $B_i$).

As it is difficult to construct finite element functions in $K_B$, we
choose to use a penalty method to approximate the constraint problem
$({\cal P})$ by a sequence $({\cal P}^\varepsilon)$ of unconstrained
problems. It reads
\begin{equation}
({\cal P}^\varepsilon)\quad\left\{
\begin{array}{l}
\hbox{Find } u^\varepsilon \hbox{ in } H^1_0(\Omega) \hbox{ and } p^\varepsilon  \hbox{ in }  L^2_0(\Omega) \hbox{ such that }\vspace{6pt}\\
\displaystyle
2\mu \int_\Omega \hbox{D}(u^\varepsilon) : \hbox{D}(\tilde u) + \frac{2}{\varepsilon} \int_B \hbox{D}(u^\varepsilon) : \hbox{D}(\tilde u) 
- \int_\Omega p^\varepsilon \nabla\cdot\tilde u
= \int_\Omega f \cdot \tilde u
,\,\,\, \forall 
 \tilde u \in H^1_0(\Omega),\vspace{6pt}\\
\displaystyle\int_\Omega q\nabla \cdot u^\varepsilon = 0 ,\,\,\,
\forall q\in L^2_0(\Omega).\nonumber
\end{array}
\right. 
\end{equation}
One can prove (see~\cite{janela2005penalty}) that, when $\varepsilon$
goes to zero, $u^\varepsilon$ tends to the solution $u$ of $({\cal
  P})$. Note that this method can also be described by considering the
rigid domain as a fluid with infinite viscosity. 

Finally, let's the time step $\dt$ be given and suppose that the time
interval $[0,T]$ is discretized using $M+1$ points $t^0=0, \ldots,t^n,\ldots,t^M=T$
with $t^{n+1}-t^n=\dt$.

For each $n$, if the rigid domain $B^n$ and the external force $f^n$
at time $t^n$ are known, we propose the algorithm \ref{alg:1} to
compute the new position of $B^{n+1}$.
\algsetup{indent=2em}
\begin{algorithm}[h!]
  \caption{First Algorithm \label{alg:1}}
    \begin{algorithmic}[1]
      \STATE Compute $(u^{n},p^{n})$ solution of $({\cal
        P}^\varepsilon)$ with $B=B^n$ and $f=f^n$,
      \STATE Compute the
      corresponding velocities of the particles: $\displaystyle
      V^n_i=\frac{1}{|B_i^n|}\int_{B_i^n} u^{n}$,
      \STATE Compute $x_i^{n+1}$, the center of $B_i^{n+1}$, using the
      identity $x_i^{n+1}=x_i^n+\dt V_i^n$. 
   \end{algorithmic}
\end{algorithm}
Note that $u^{n}$, $p^{n}$, $V^n$, $B^{n+1}$ and $x_i^{n+1}$ depend on
the penalty parameter $\varepsilon$. However, the superscript
$\varepsilon$ have been skipped for the sake of readability.

\subsection{Extra constraints and final algorithm}
\label{sec:extra}

Two problems may occur when we solve the vesicle problem by using the
previous algorithm \ref{alg:1}. First, nothing is done to avoid overlaps between
the particles and to keep the neighbouring particles of the membrane
sticking together. Then, although the velocity is supposed to be
divergence-free, numerical approximation can make the area of the
vesicle change during the computation and one has to pay attention to
this point when running long-time simulation.

In order to ensure, at each time step, that particles do not overlap
and that two neighbouring particles are stuck, we add a projection
step to the previous algorithm. We project the velocities of the
particles, computed from the penalized problem $({\cal
  P}^\varepsilon)$ onto a set of admissible velocities.

To choose this set, we generalize the method
proposed in~\cite{maury2006collisions} to deal with inelastic
contacts. More precisely, let $D_{ij}(x^n)$ be the distance between particles $i$ and
$j$ at time $n$ and $G_{ij}(x^n)$ its gradient. We first linearize the
distance $D_{ij}(x^{n+1})$ as follow
\begin{equation}
  \label{eq:6}
  D_{ij}(x^{n+1})=D_{ij}(x^{n}+\dt V^n)\approx D_{ij}(x^{n})+\dt G_{ij}(x^n)\cdot V^n,  
\end{equation}
then, we chose this discrete set of admissible velocities at time $n$:
\begin{equation}
  \label{eq:7}
  K_c^n=\left\{
    \begin{array}{ll}
      \displaystyle V\in\mathbb{R}^{2N},\,\,\, & \displaystyle D_{ij}(x^n) +\dt G_{ij}(x^n)\cdot V\geq 0\quad \forall i<j \vspace{6pt}\\
      & \displaystyle D_{i,i+1}(x^n) +\dt G_{i,i+1}(x^n)\cdot V \leq 0\quad \forall i\in[1\ldots N-1] \vspace{6pt}\\
      & \displaystyle D_{N,1}(x^n) +\dt G_{N,1}(x^n)\cdot V\leq 0
    \end{array}
  \right\}.
\end{equation}

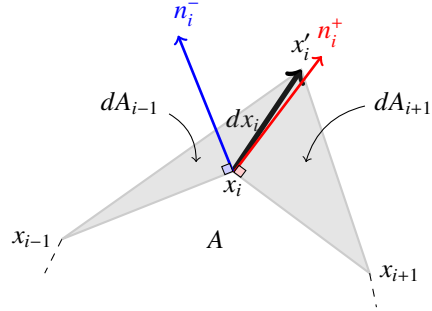
\begin{figure}[!h]
  \centering
   \begin{tikzpicture}[scale=0.45]
  \tkzDefPoint[label=below:$x_i$](5,4){xi}
  \tkzDefPoint[label=left:$x_{i-1}$](0,2){xim1}
  \tkzDefPoint[label=right:$x_{i+1}$](9,1){xip1}
  \tkzDefPoint[label=above:$x_i^\prime$](7,7){xid}
  
  \path[->] [black,thick,line width=2pt] (xi) edge (xid);
  \draw[black]  node (dx) at (5.3,5.5) {${dx}_i$};
  
  \draw[black,thick,fill=gray, opacity=0.2] (xim1) -- (xid) -- (xip1) -- (xi) -- (xim1);
  
  \coordinate (xim1p) at (-0.5,1);
  \coordinate (xip1p) at (9.2,0);
    
  \path[dashed] (xip1) edge (xip1p);
  \path[dashed] (xim1) edge (xim1p);
      
  \tkzDefPoint(7.6,7.4){C}
  \tkzLabelPoint[red,label={\color{red}${n}_i^{+}$}](C){}; 
  \tkzDefPointBy[projection=onto xi--xip1](C)  \tkzGetPoint{H}
  \path[->] [red, thick, line width=1pt] (H) edge (C);
  \tkzMarkRightAngle[fill=red!20,size=.25](C,H,xip1)
    
  \tkzDefPoint(3.4,8){D}
  \tkzLabelPoint[blue,label={\color{blue}${n}_i^{-}$}](D){}; 
  \tkzDefPointBy[projection=onto xim1--xi](D)  \tkzGetPoint{G}
  \path[->] [blue, thick, line width=1pt] (G) edge (D);
  \tkzMarkRightAngle[fill=blue!20,size=.25](D,G,xim1)
    
  \node (a) at (10,6){$dA_{i+1}$};
  \node (b) at (7,4) {};
  \path[->] (a) edge [bend right]  (b);
    
  \node (a1) at (2,6){$dA_{i-1}$};
  \node (b1) at (4,4) {};
  \path[->] (a1) edge [bend left]  (b1);

  \node (a2) at (4.5,2){$A$};
\end{tikzpicture}
  \caption{Area variation induced by the displacement $dx_i$ of particle $i$.}
  \label{fig:varvol}
\end{figure}

A second projection step is added in order to deal with the constant
area constraint. To do so, let's assume that the area of the vesicle
is approximated by the area of the polygon defined by the center of
the particles. Then, we compute the area variation $dA$ induced by the
displacement $dx_i$ of particle $i$. We denote by $n_i^+$ and $n_i^-$
the two outer unit normals to the polygon at point $x_i$ (see
figure~\ref{fig:varvol}) and we define
$n_i=\frac{1}{2}(|x_{i+1}-x_i|n_i^+ +|x_{i-1}-x_i|n_i^-)$ (note that
indices has to be adapted for $i=1$ and $i=N$).
Then, the considered area variation is given by
\[
\begin{array}{ll}
\displaystyle dA&\displaystyle =dA_{i+1}+dA_{i-1}\vspace{6pt}\\
 & \displaystyle=\frac{|x_{i+1}-x_i|dx_i\cdot
  n_i^+}{2}+\frac{|x_{i-1}-x_i|dx_i\cdot
  n_i^-}{2}\vspace{6pt}\\
 & \displaystyle
=n_i\cdot dx_i.
\end{array}
\]
Obviously, in the case where all particles move we obtain
\[\displaystyle dA = \sum_{i=1}^N n_i\cdot dx_i,\]
and the constraint at time $n$ is written
\[A^0=A^{n+1}=A^n+dA=A^n+\dt \sum_{i=1}^N n_i\cdot V_i^n.\]

Finally, the space of admissible velocities for the area constraint at time $n$ is given by
\[
K_v^n=\left\{
V\in\mathbb{R}^{2N},\,\,\, A^0=A^n+\dt \sum_{i=1}^N n_i\cdot V_i\right\}
\]
and the global algorithm is given by algorithm \ref{alg:final},
\algsetup{indent=2em}
\begin{algorithm}[h!]
  \caption{Final Algorithm \label{alg:final}}
  \begin{algorithmic}[1]
    \STATE Compute $(u^{n},p^{n})$ solution of $({\cal P}^\varepsilon)$
    with $B=B^n$ and $f=f^n$,
    \STATE Compute the corresponding velocities of the particles:
    $\displaystyle \tilde V^n_i=\frac{1}{|B_i^n|}\int_{B_i^n} u^{n}$,
    \STATE Deal with contacts by projecting $\tilde V^n$ on  $K_c^n$ :
    $\hat V^n = \Pi_{K_c^n} \tilde V^n$,
    \STATE Deal with the area constraint by projecting  $\hat V^n$ on
    $K_v^n$ : $V^n = \Pi_{K_v^n} \hat V^n$,
    \STATE Compute the new position of $B^{n+1}$: $x_i^{n+1}=x_i^n+\dt V_i^n$,
  \end{algorithmic}
\end{algorithm}
where $\Pi_K$ denotes the projection onto $K$ and is performed using a Uzawa algorithm.

\section{Dimensionless numbers}
\label{sec:dimens-numb}

To validate our model by comparing its results with those from the
literature, we need to define some relevant dimensionless parameters which
control the problem.
\begin{description}
\item[The reduced Area] $\alpha$ is defined as the ratio between the
  vesicle's area and the area of a circle having the same perimeter.
  \begin{equation}
    \label{eq:8}
    \alpha=\frac{A}{\pi R_0^2},
  \end{equation}
where $A$ is the area of the vesicle and $R_0$ its effective radius
given by $\frac{P}{2\pi}$ ($P$ being the perimeter of the vesicle).

The reduced area $\alpha$ measure the amount of deflation of the
vesicle. We will see in section \ref{sec:num-results} that this
parameter controls the equilibrium shape of the vesicle.
\item[The Viscosity Contrast] $\lambda$ is defined as the ratio
  between the viscosities of the inner and the outer fluids. It reads
  \begin{equation}
    \label{eq:9}
    \lambda=\frac{\mu_{in}}{\mu_{out}}.
  \end{equation}
  This
  parameter monitors the dynamic of the vesicle (see section
  \ref{sec:num-results}).
\item[The confinement] $\tau$ is given by 
  \begin{equation}
    \label{eq:10}
    \tau=\frac{2R_0}{l}    
  \end{equation}
  where $l$ is the height of the chanel in which the vesicle evolves.
\item[The Capilary number] $C_a$ (see \ref{sec:appendixCaN})
  which can be defined as
  \begin{equation}
    \label{eq:11}
    C_a=\frac{\mu_{out} R_0^3 \dot\gamma}{2k_a r}
  \end{equation}
  where $\dot\gamma$ stands for the shear rate and $r$ for the radius of the particles in our model.

  The Capillary number can be seen as the ratio between the
  characteristic time of the shear and a characteristic time related
  to the bending force. This number measures the amount of deformation of
  the vesicle. Indeed, for high values of $C_a$, the vesicle is more
  deformable due to the hydrodynamic forces, while for low $C_a$, the
  bending energy requires a higher cost to change the curvature of
  the vesicle.
\end{description}

\section{Numerical Results}
\label{sec:num-results}

We present in this section some numerical results obtained by using
algorithm \ref{alg:final}. 
The penalized Stokes problem $({\cal P}^\varepsilon)$ is discretized 
using first order Finite Element where the discrete space for the pressure
is based on Lagrange polynomials $P^1$ and the
discrete space for the velocity field is based on
$P^1bubble$ (which is an enrichment of $P^1$ to satisfy the discrete
inf-sup condition).

For each of the following simulations, the time step is $\dt=5.10^{-3}$,
the reference length of the springs is $\ell_0=2r$ where $r=1.5$ is the
radius of the rigid particles, the penalty parameter is $\varepsilon=5.10^{-3}$ and
the shear rate is $\dot\gamma=1$.

The other parameters are, for the different sections:

\begin{tabular}{c|cccccccc}
Section & $N$ & $k_a$ & $k_{rp}$ & $l$ & $L$ & $\mu_{in}$ &
$\mu_{out}$\\\hline
Equilibrium shapes & 42 & 200 & 0.25 & 150 & 150 & 1 & 1\\
TT & 50 & 600 & 0.5 & 242 & 300 & 1 & 1 \\
TB & 50 & 600 & 0.5 & 242 & 300 & 1 & - \\
VB & 50 & 600 & 0.5 & 242 & 300 & 1 & - 
\end{tabular}

\subsection{Equilibrium shapes}

An essential step for the validation of any vesicle model is to
retrieve the equilibrium shape of a vesicle in a fluid at rest.  More
precisely, if we consider a vesicle in a shape which is not the one
minimizing its energy, the curvature force makes the membrane move as
the fluid around it. This dynamic stops when the vesicle reaches its
equilibrium state in which the final shape of the vesicle depends on
its reduced area.

In our simulations the initial distribution of the rigid particles
(that form the vesicle membrane) is done by placing them in an ellipse.
Then, the Stokes problem is solved in a rectangular domain containing
the vesicle with homogeneous Dirichlet boundary conditions.

\begin{figure}[hbtp]
  \centering
  \begin{tikzpicture}
  \begin{axis}[
    axis equal=true,
    grid=both,minor tick num=3, xlabel=x,ylabel=y,
    ]
    \addplot+[smooth,blue!80!black,mark=0,solid,thick]table[x=X,y=Y]{RA042Positions00000};
    \addlegendentry{$t=0$}
    \addplot+[smooth,magenta!80!black,mark=0,dashed,thick]table[x=X,y=Y]{RA042Positions01000};
    \addlegendentry{$t=5$}
    \addplot+[smooth,green!80!black,mark=0,densely dashed,thick]table[x=X,y=Y]{RA042Positions02000};
    \addlegendentry{$t=10$}
    \addplot+[smooth,brown!80!black,mark=0,solid,thick]table[x=X,y=Y]{RA042Positions04740};
    \addlegendentry{$t=23.7$}
  \end{axis}
\end{tikzpicture}
  \caption{Equlibrium shapes. Shapes versus $t$ for $\alpha=0.42$.}
  \label{fig:equilibre_shape_vs_t}
\end{figure}
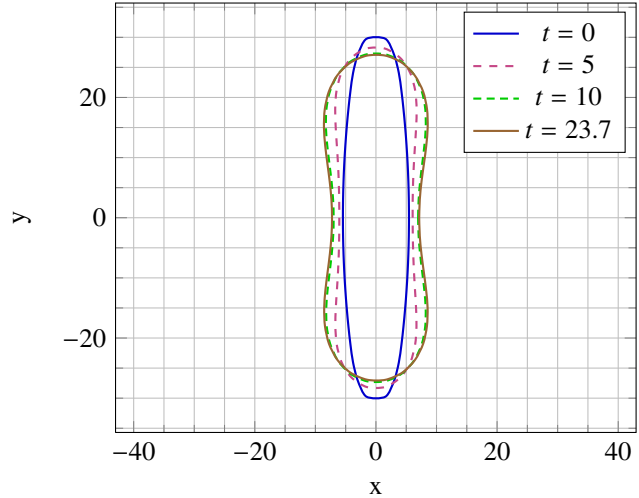

Figure~\ref{fig:equilibre_shape_vs_t} shows the evolution of a vesicle
suspended in a fluid at rest by plotting the position of each rigid
particle center. One can see that the vesicle
reaches its equilibrium shape which is biconcave in the case of small
reduced area. Note that this form is typical of red blood cells

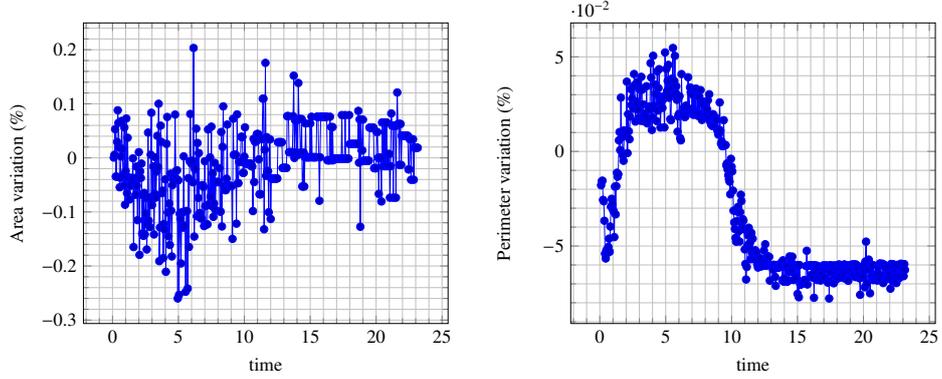
\begin{figure}[hbtp]
  \centering
  \begin{tikzpicture}[scale=0.7]
    \begin{axis}[xlabel=time,ylabel=Area variation (\%), 
      grid=both, ,minor tick num=4,
      ]

      \addplot
      table[x=t,y=V]{VolPerim_Eq04.conf0.27.eps5e3-INIT10.txt}; 
      
    \end{axis}
  \end{tikzpicture}  \quad
  \begin{tikzpicture}[scale=0.7]
    \begin{axis}[xlabel=time,ylabel=Perimeter variation (\%), 
      grid=both, ,minor tick num=4,
      ]

      \addplot
      table[x=t,y=P]{VolPerim_Eq04.conf0.27.eps5e3-INIT10.txt}; 
      
    \end{axis}
  \end{tikzpicture}  
  \caption{Equilibrium shapes. Area and perimeter versus $t$ for $\alpha=0.42$.}
  \label{fig:equilibre_VolPerim_vs_t}
\end{figure}

It is important to check that the area and the perimeter are well
conserved.  During this simulation, the variation of the area ranges
from $-0.3\%$ to $0.2\%$ and the variation of the perimeter from
$-0.05\%$ to $0.05\%$ (see figure~\ref{fig:equilibre_VolPerim_vs_t}).

\begin{figure}[h]
  \centering
   \begin{tikzpicture}
  \begin{axis}[
    axis equal=true,
    grid=both,minor tick num=3, xlabel=x,ylabel=y,
    legend style={legend cell align=left, legend pos=outer north
      east}
    ]
    \addplot+[smooth,blue!80!black,mark=0,solid,thick]table[x=X,y=Y]{RA042Positions04740};
    \addlegendentry{$\alpha=0.42$}
    \addplot+[smooth,magenta!80!black,mark=0,dashed,thick]table[x=X,y=Y]{RA051Positions04590};
    \addlegendentry{$\alpha=0.51$}
    \addplot+[smooth,gray!80!black,mark=0,solid,thick]table[x=X,y=Y]{RA061Positions04800};
    \addlegendentry{$\alpha=0.61$}
    \addplot+[smooth,green!80!black,mark=0,densely dashed,thick]table[x=X,y=Y]{RA072Positions03500};
    \addlegendentry{$\alpha=0.72$}
    \addplot+[smooth,brown!80!black,mark=0,solid,thick]table[x=X,y=Y]{RA081Positions01000};
    \addlegendentry{$\alpha=0.81$}
    \addplot+[smooth,red!80!black,mark=0,densely dashed,thick]table[x=X,y=Y]{RA09Positions01000};
    \addlegendentry{$\alpha=0.9$}   
    \addplot+[smooth,black!80!black,mark=0,solid,thick]table[x=X,y=Y]{RA1Positions00290};
    \addlegendentry{$\alpha=1$}
  \end{axis}
\end{tikzpicture}
  \caption{Equilibrium shapes for different values of $\alpha$.}
  \label{fig:equilibre}
\end{figure}
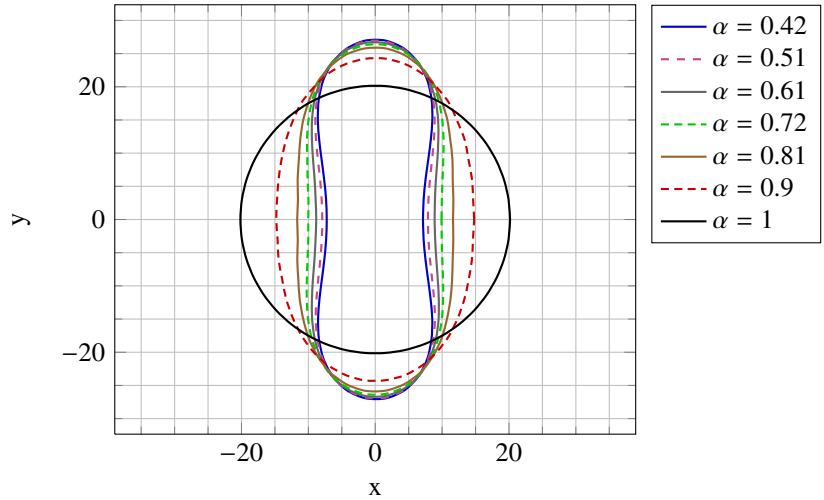

Figure \ref{fig:equilibre} shows the computed equilibrium shapes of
vesicles with different reduced area. As it is well known, the
equilibre shape goes from circular one when $\alpha=1$ to biconcave one when
$\alpha$ is small enough (about $0.6$).

\begin{figure}[hp]
  \centering
  \begin{tikzpicture}[mark size={3.6},scale=0.6]
  \begin{axis}[
    xmin=-38, xmax=38,ymin=-32, ymax=32,axis equal=true,
    grid=both,minor tick num=3, 
    title={$\alpha=0.42$},
    ]
    \addplot+[smooth,black!80!magenta,thick,only marks,mark=o]
    table[x=X,y=Y]{RA042Positions04740};
    \addlegendentry{Present work}
  \end{axis}
\end{tikzpicture}
\begin{tikzpicture}[mark size={3.6},scale=0.6]
  \begin{axis}[
    xmin=-38, xmax=38,ymin=-32, ymax=32,axis equal=true,
    grid=both,minor tick num=3, 
    title={$\alpha=0.51$},
    ]  
    \addplot+[smooth,black!80!magenta,thick,only marks,mark=o]
    table[x=X,y=Y]{RA051Positions04590};
    \addlegendentry{Present work}
  \end{axis}
\end{tikzpicture}
\begin{tikzpicture}[mark size={3.6},scale=0.6]
  \begin{axis}[
    xmin=-38, xmax=38,ymin=-32, ymax=32,axis equal=true,
    grid=both,minor tick num=3, 
    title={$\alpha=0.61$},
    ]  
    \addplot+[smooth,black!80!magenta,thick,only marks,mark=o]
    table[x=X,y=Y]{RA061Positions04800};
    \addlegendentry{Present work}
    \addplot+[smooth,red!80!black,mark=0,solid,thick]table[x=X,y=Y]{LBM_060.dat};
    \addlegendentry{\cite{Kaoui2011}}      
  \end{axis}
\end{tikzpicture}
\begin{tikzpicture}[mark size={3.6},scale=0.6]
  \begin{axis}[
    xmin=-38, xmax=38,ymin=-32, ymax=32,axis equal=true,
    grid=both,minor tick num=3, 
    title={$\alpha=0.72$},
    ]  
    \addplot+[smooth,black!80!magenta,thick,only marks,mark=o]
    table[x=X,y=Y]{RA072Positions03500};
    \addlegendentry{Present work}  
    \addplot+[smooth,red!80!black,mark=0,solid,thick]table[x=X,y=Y]{LBM_070.dat};
    \addlegendentry{\cite{Kaoui2011}}      
  \end{axis}
\end{tikzpicture}
\begin{tikzpicture}[mark size={3.6},scale=0.6]
  \begin{axis}[
    xmin=-38, xmax=38,ymin=-32, ymax=32,axis equal=true,
    grid=both,minor tick num=3, 
    title={$\alpha=0.81$},
    ]  
    \addplot+[smooth,black!80!magenta,thick,only marks,mark=o]
    table[x=X,y=Y]{RA081Positions01000};
    \addlegendentry{Present work}  
    \addplot+[smooth,red!80!black,mark=0,solid,thick]table[x=X,y=Y]{LBM_080.dat};
    \addlegendentry{\cite{Kaoui2011}}      
  \end{axis}
\end{tikzpicture}
\begin{tikzpicture}[mark size={3.6},scale=0.6]
  \begin{axis}[
    xmin=-38, xmax=38,ymin=-32, ymax=32,axis equal=true,
    grid=both,minor tick num=3, 
    title={$\alpha=0.9$},
    ]  
    \addplot+[smooth,black!80!magenta,thick,only marks,mark=o]
    table[x=X,y=Y]{RA09Positions01000};
    \addlegendentry{Present work}  
    \addplot+[smooth,red!80!black,mark=0,solid,thick]table[x=X,y=Y]{LBM_090.dat};
    \addlegendentry{\cite{Kaoui2011}}      
  \end{axis}
\end{tikzpicture}
\begin{tikzpicture}[mark size={3.6},scale=0.6]
  \begin{axis}[
    xmin=-38, xmax=38,ymin=-32, ymax=32,axis equal=true,
    grid=both,minor tick num=3, 
    title={$\alpha=1$},
    ]  
    \addplot+[smooth,black!80!magenta,thick,only marks,mark=o]
    table[x=X,y=Y]{RA1Positions00290};
    \addlegendentry{Present work}  
    \addplot+[smooth,red!80!black,mark=0,solid,thick]table[x=X,y=Y]{LBM_100.dat};
    \addlegendentry{\cite{Kaoui2011}}      
  \end{axis}
\end{tikzpicture}
  \caption{Equilibrium shapes. Comparison with results from Kaoui et al. \cite{Kaoui2011}}
  \label{fig:ComparaisonEqShape}
\end{figure}
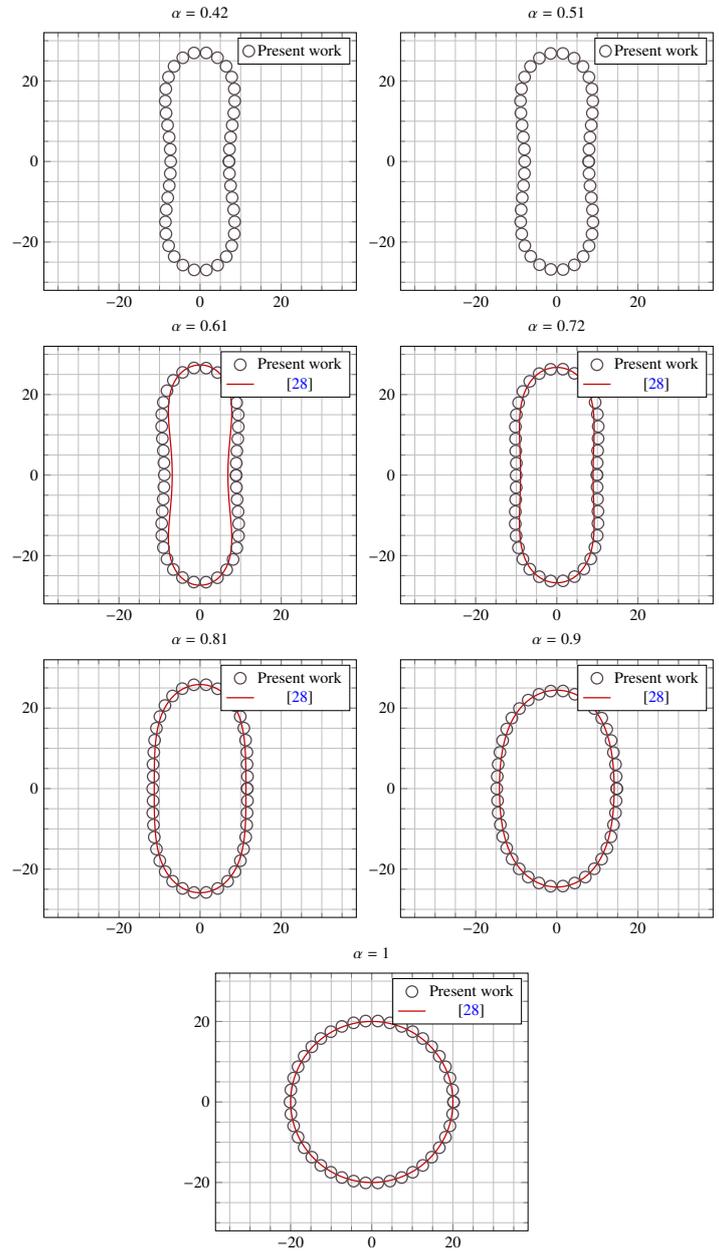

Finally, we compare our results to those presented in \cite{Kaoui2011}
(in this paper equilibrium shapes are computed using two numerical
methods : Boundary Integral and Lattice Boltzman). Figure
\ref{fig:ComparaisonEqShape} shows at the same times our results and
those from \cite{Kaoui2011} for different value of the reduced area
parameter. One can notice that there is a good agreement between these
results. Nevertheless, we see small differences in the cases where
$\alpha$ is small and we believe this is due to the thickness of our
membrane.

\subsection{``Tank-Treading'' Regime (TT)}
\label{sec:tank-treabing-regime}

\begin{figure}[!ht]
  \centering
     \begin{tikzpicture}[scale=0.7]
    \begin{axis}[xlabel=Reduced Area
      $\alpha$,ylabel=${\displaystyle\frac{\theta}{\pi}}$, 
      title={Tank-Treading $\lambda=1$}, grid=both, ,minor tick num=4,
      legend cell align=left, legend pos=north west
      ]

      \addplot table[x=RA,y=thetaRpi]{newAnglesTT.conf0.2.dat}; 
      \addlegendentry{Present work $\tau=0.2$}
      
      \addplot table[x=RA,y=BeaucourtRR]{beaucourt.dat}; 
      \addlegendentry{\cite{beaucourt2004steady} $\tau=0$}


    \end{axis}
  \end{tikzpicture}  
  \caption{Tank treading angles. Comparison between our results (in
    the case of small confinement $\tau=0.2$ ) and
    those from \cite{beaucourt2004steady} (in unbounded domain).}
  \label{fig:TT1Bis}
\end{figure}
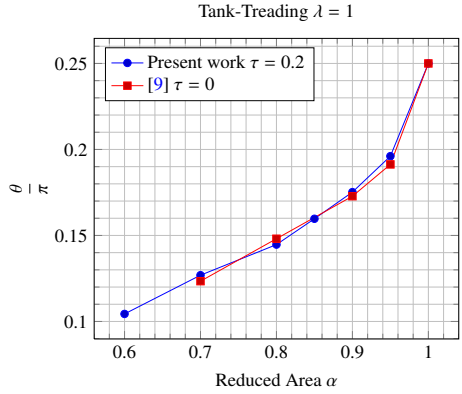

We are interested in this section to the dynamic of a vesicle under
linear shear flow. It is well known that if $\lambda$, the ratio
between the inner and the outer fluid viscosities, is smaller than a
certain critical value (around $5$ in unbounded domain
\cite{beaucourt2004steady}), the vesicle takes a steady angle with
respect to the horizontal axis and its membrane starts to rotate like
a chain of a tank (this motion is called Tank-Treading). To
validate our model in this regime, we refer to the results given in
\cite{beaucourt2004steady}. So, we consider vesicles of different
reduced area subject to linear shear flow in a rectangular domain and the
confinement parameter $\tau$ is chosen to be small (around $0.2$) to
minimize the effect of walls. In figure \ref{fig:TT1Bis} we plot the
steady angle for different reduced area and we compare our results
with those from \cite{beaucourt2004steady}.





\begin{figure}[h]
  \centering
  \subfigure[$t=5.10^{-3}$]{\includegraphics[scale=0.15]{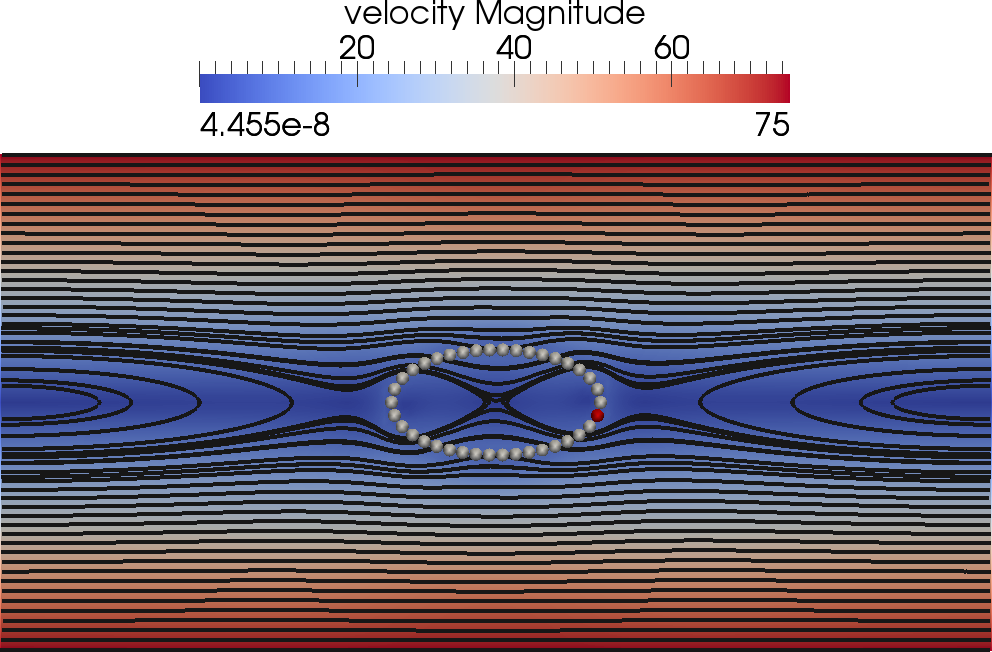}}
  \subfigure[$t=0.5$]{\includegraphics[scale=0.15]{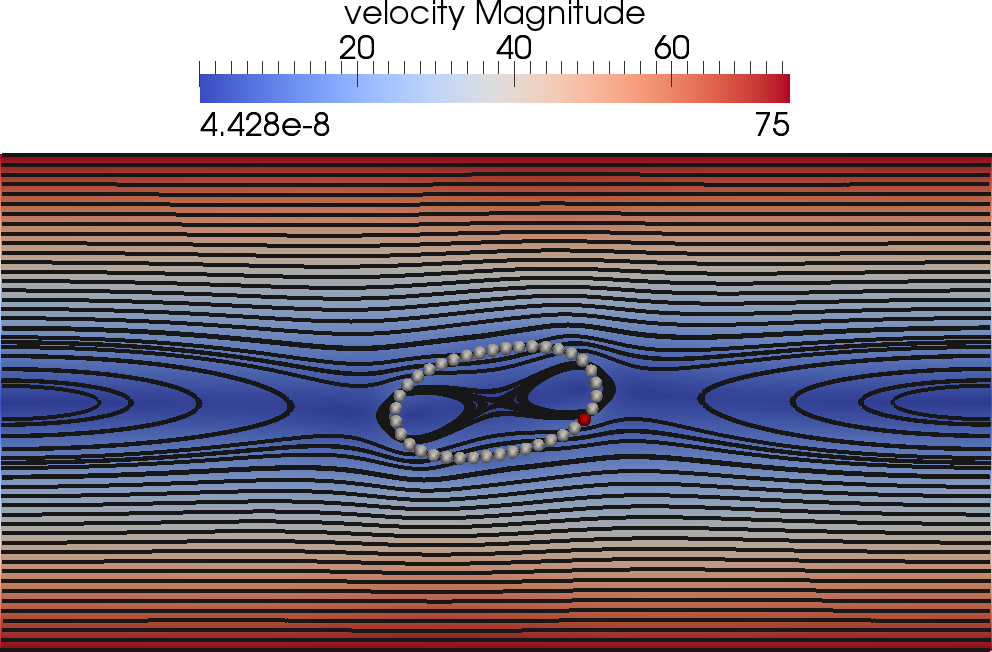}}
  \subfigure[$t=1.5$]{\includegraphics[scale=0.15]{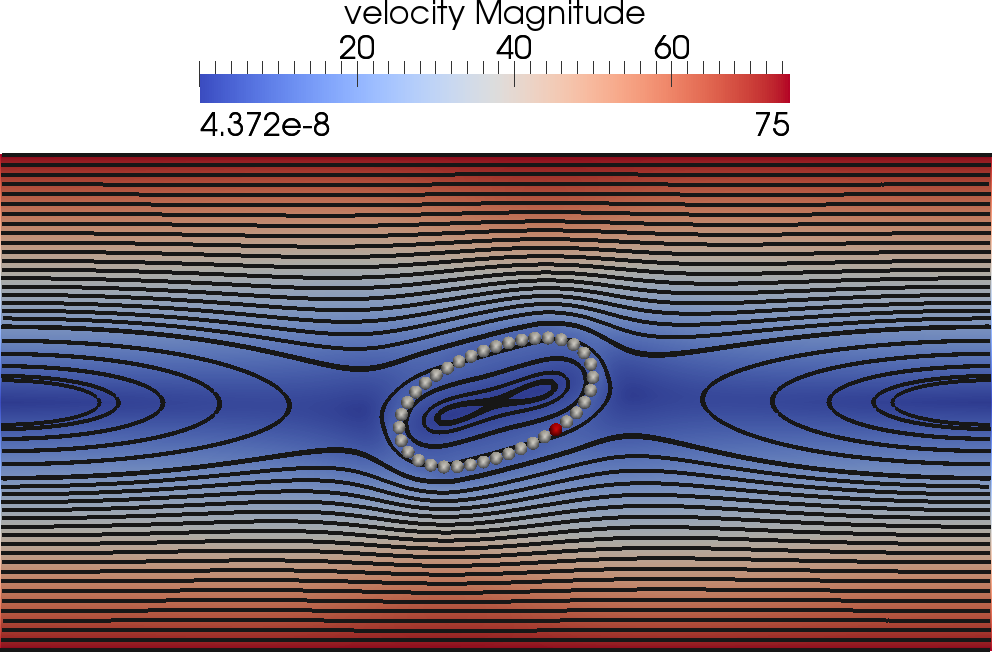}}
  \subfigure[$t=2.95$]{\includegraphics[scale=0.15]{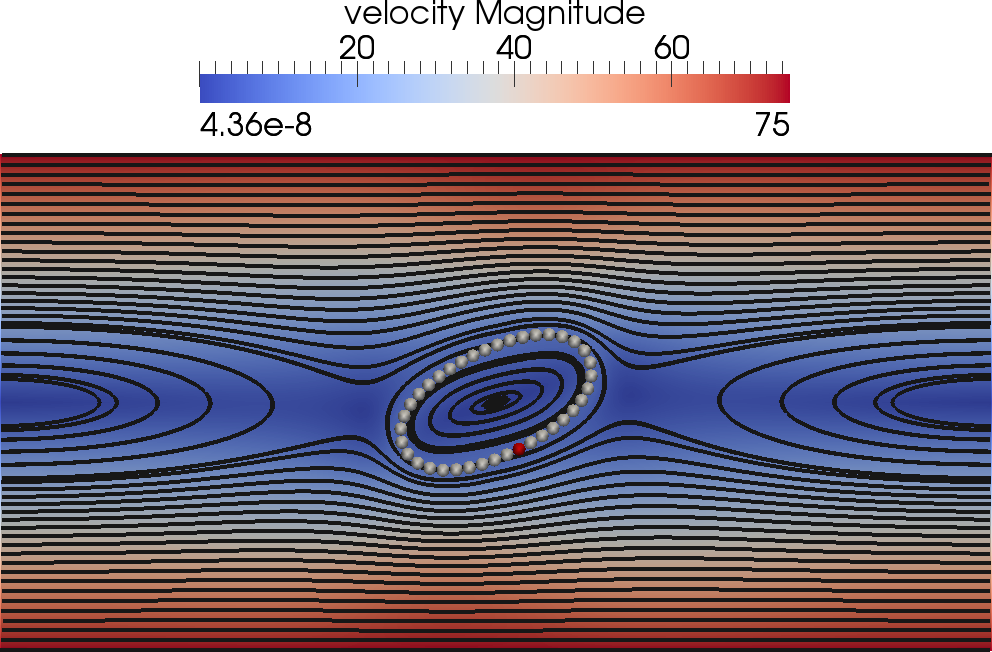}}
  \subfigure[$t=15.7$]{\includegraphics[scale=0.15]{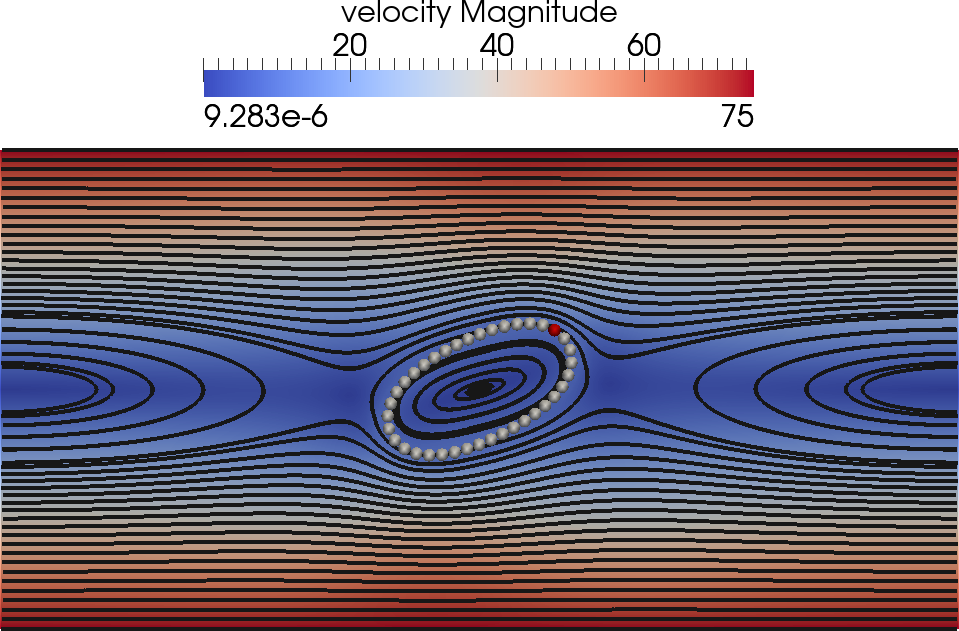}}
  \caption{Streamlines of the velocity field and vesicle position in
    Tank-Treading regime at different time. $\alpha=0.85$,
    $\tau=0.32$, $N=38$ and $\lambda=1$.
  }
  \label{fig:streamTT}
\end{figure}

Finally we plot in figure \ref{fig:streamTT} the streamlines of the
velocity field, its magnitude and the position of the vesicle at
different times. In this simulation the reduced area is $\alpha=0.85$,
the confinement parameter is  $\tau=0.32$, the viscosity contrast
$\lambda=1$ and the number of rigid particles is $N=38$.

\subsection{Tumbling Regime (TB)}
\label{sec:tumbling-regime}

When the viscosity contrast $\lambda$ is above a certain critical
value $\lambda_c$, a vesicle subject to linear shear flow starts to
rotate and tends to follow a solid rotation. This motion is called
tumbling and is also typical of red blood cells.

\begin{figure}[h]
  \centering
  \subfigure[$t=5.10^{-3}$]{\includegraphics[scale=0.15]{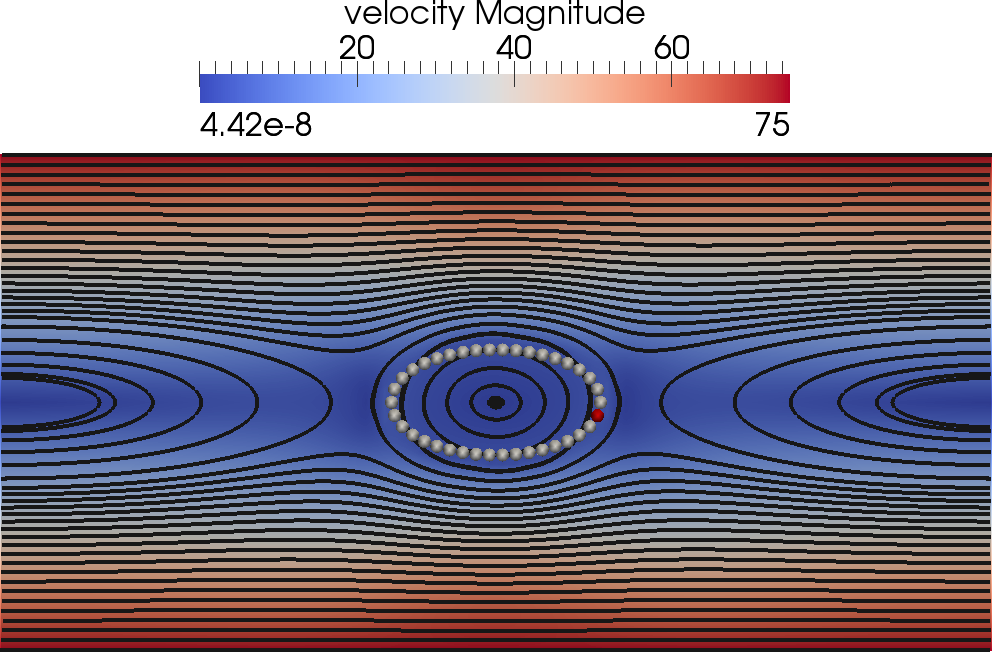}}
  \subfigure[$t=2.7$]{\includegraphics[scale=0.15]{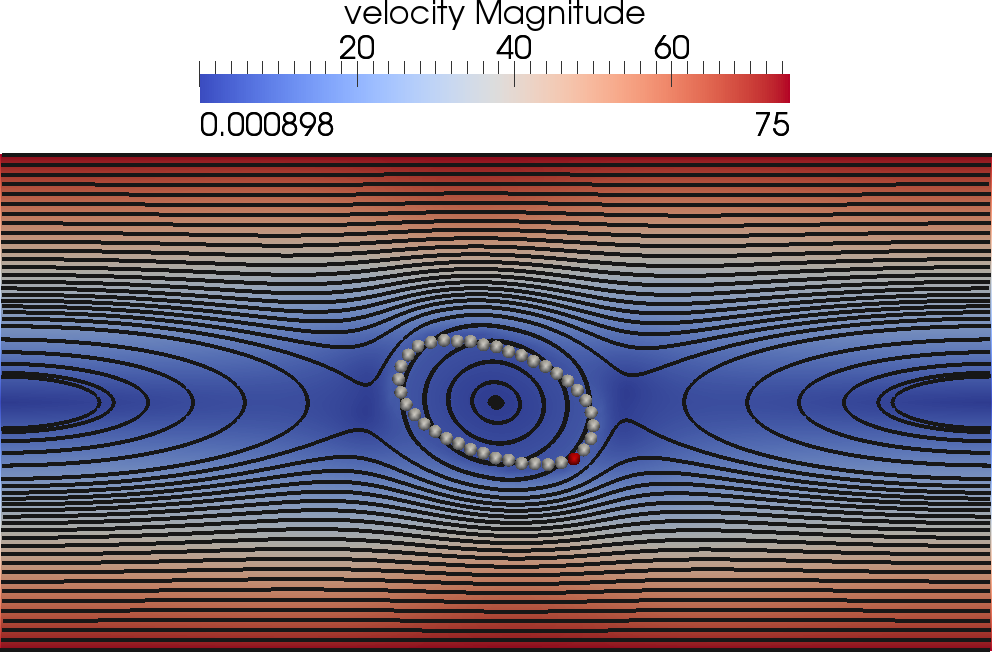}}
  \subfigure[$t=5.35$]{\includegraphics[scale=0.15]{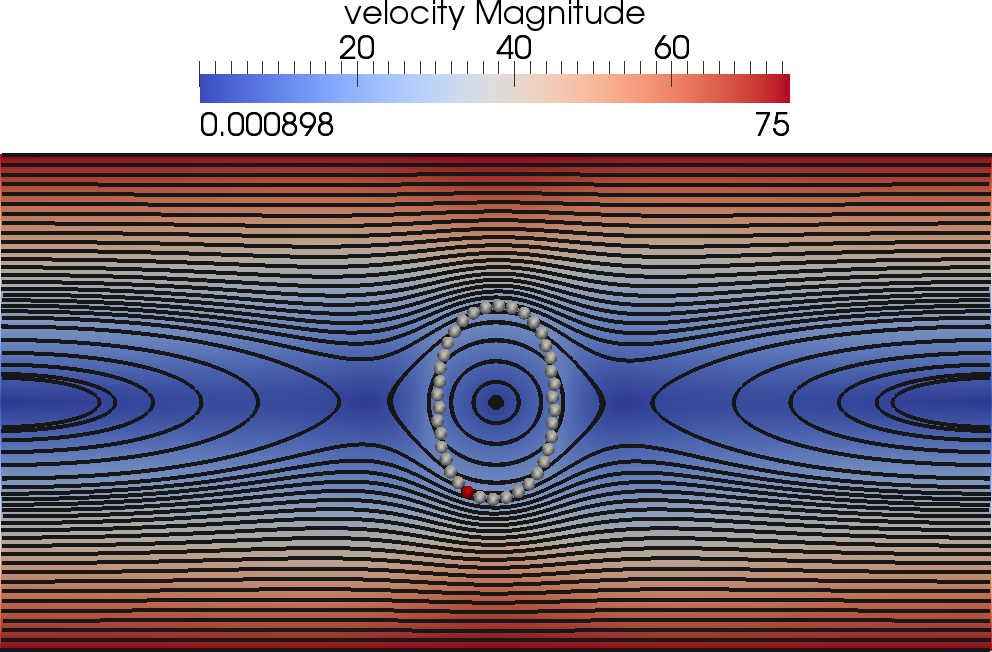}}
  \subfigure[$t=8.05$]{\includegraphics[scale=0.15]{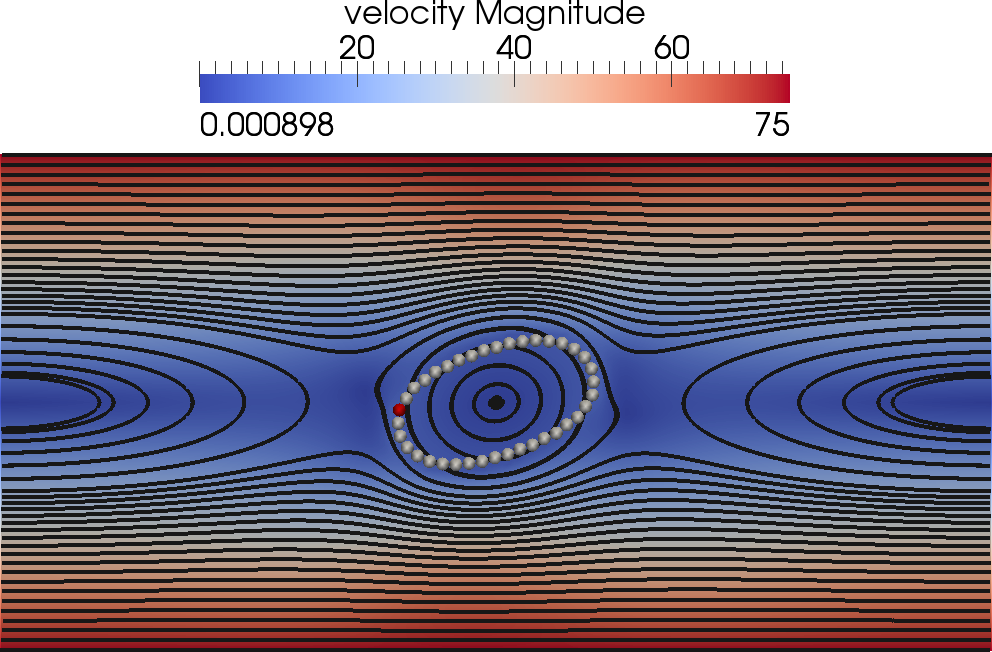}}
  \subfigure[$t=10.05$]{\includegraphics[scale=0.15]{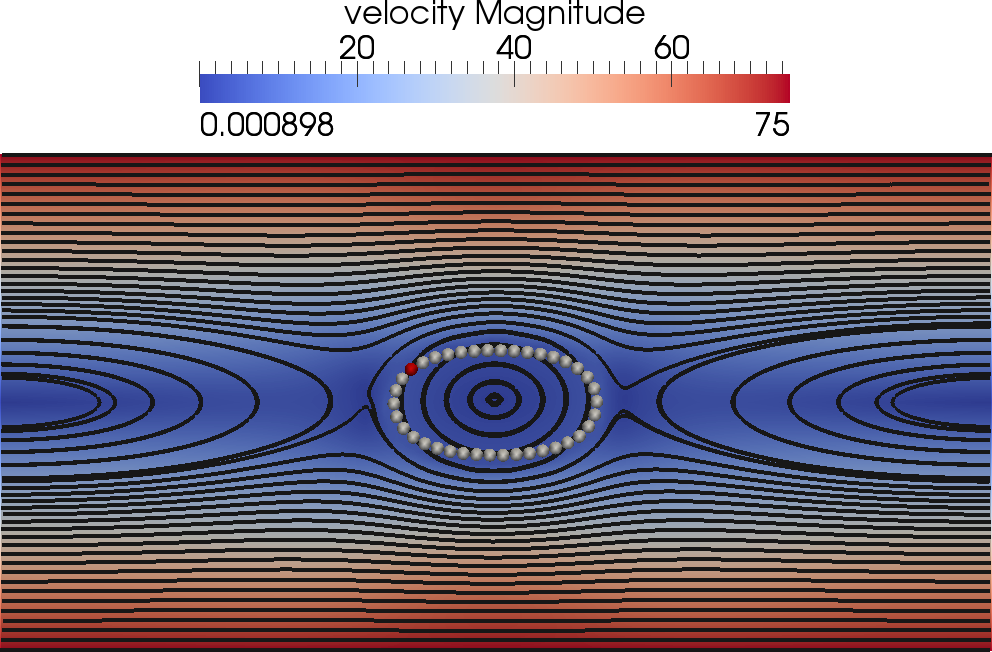}}
  \caption{Streamlines of the velocity field and vesicle position in
    Tumbling regime at different time. $\alpha=0.85$, $\tau=0.32$,
    $N=38$ and $\lambda=20$.}
  \label{fig:streamTB}
\end{figure}

Figure \ref{fig:streamTB} shows the streamlines of the
velocity field, its magnitude and the position of the vesicle at
different times. The initial time corresponds to the first time step
iteration and the final one corresponds to the time during which the
vesicle made half a turn.
In this simulation we keep the same parameters as in the case of TT
simulation except the viscosity contrast $\lambda=20$.

\begin{figure}
  \centering
  \includegraphics[scale=0.6]{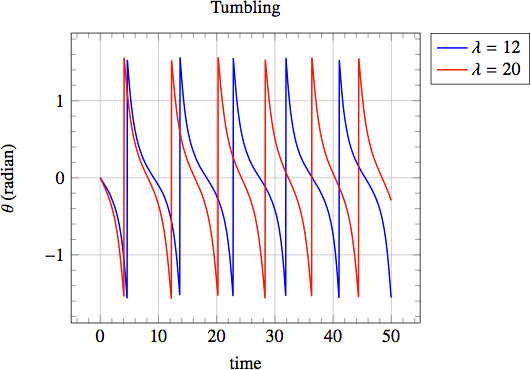}
  \caption{Angles versus time in tumbling regime. $\alpha=0.85$ and $\tau=0.2$.}
  \label{fig:TB12TB20}
\end{figure}

If we look at the particle marked in red, we note that the membrane
has also turned slightly on itself. The vesicle rotation is not yet that of a solid. See
\cite{DOYEUX:2012:HAL-00665481:1} in which the authors showed that increasing the viscosity
ratio increases the rotation frequency of the vesicle. This frequency
reaches a steady value which corresponds to the rotation of a solid
object. In this work, we show in figure \ref{fig:TB12TB20} the
angles of a vesicle in tumbling regime versus time for two values of
$\lambda$ ($12$ and $20$). We can see easily that the frequency
rotation when $\lambda=20$ is greater than in the case $\lambda=12$.

\subsection{Transition between Tank-Treading and
  Tumbling.  The ``Vacillating -Breathing'' Regime (VB)}
\label{sec:trans-betw-tt}

In addition to the two previous regimes TT and TB, it has been shown
(see \cite{PhysRevLett.96.028104} and \cite{PhysRevLett.98.128103})
that in a shear flow, a vesicle may exhibit another kind of motion
called \textit{``Vacillating-Breathing''} in
\cite{PhysRevLett.96.028104} and \textit{``Swinging''} in
\cite{PhysRevLett.98.128103}. This regime occurs when the vesicle's
membrane is fairly deformable (for a high enough capillary number
$C_a$) and precedes the TB regime when increasing the viscosity ratio
$\lambda$.  Thus, the vesicle shape undergoes large deformations, the
long main axis of the vesicle performs oscillations around the flow
direction.  More precisely, \textit{(i)} when the vesicle is
perpendicular to the shear axis, it is being deformed till it attains
a quasi circular shape, and then \textit{(ii)} when it tends to be aligned with the shear
axis it is elongated like in the case of TT regime.

In addition to theoretical and experimental results presented
respectively in \cite{PhysRevLett.96.028104} and
\cite{PhysRevLett.98.128103} to describe this motion, we can find some
numerical studies highlighting the VB regime but they are still
infrequent. See \cite{PhysRevE.83.031921} for 3D simulation using
Boundary Integral Method and \cite{PhysRevE.80.011901} for 2D
simulation by adding thermal fluctuation. To our knowledge, our work
is the first one that presents 2D numerical simulation of VB regime
without adding thermal fluctuation. Moreover, in
\cite{ghigliotti2010rheology} and \cite{PhysRevE.80.011901} the
authors argue that there is no support to the existence of VB regime
in 2D unless the addition of high-order undulation modes.

\begin{figure}[h]
  \centering
  \subfigure[$t=5.10^{-3}$]{\includegraphics[scale=0.15]{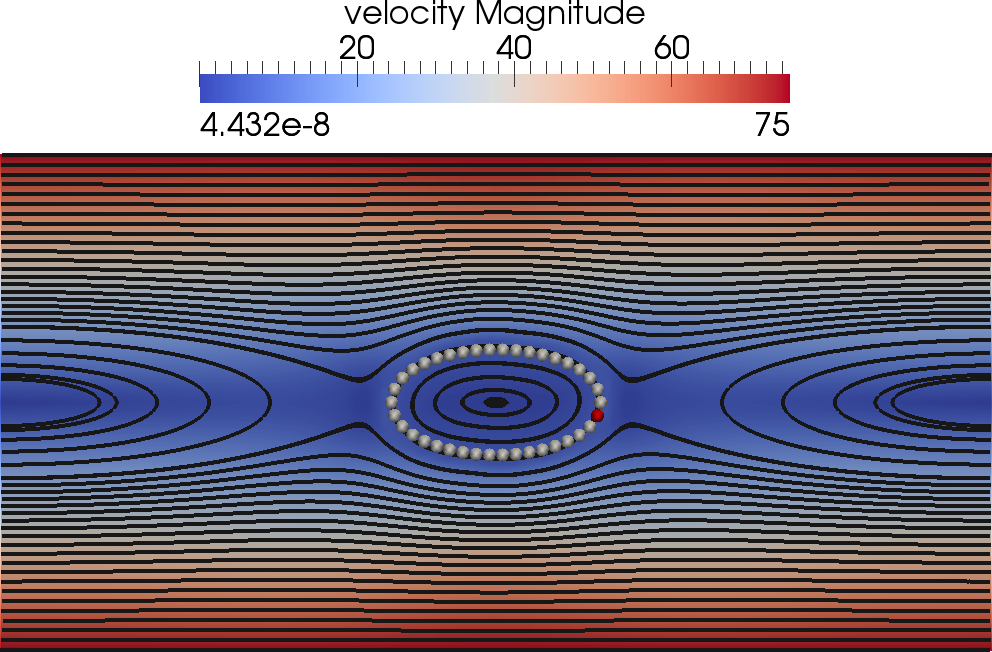}}
  \subfigure[$t=6.85$]{\includegraphics[scale=0.15]{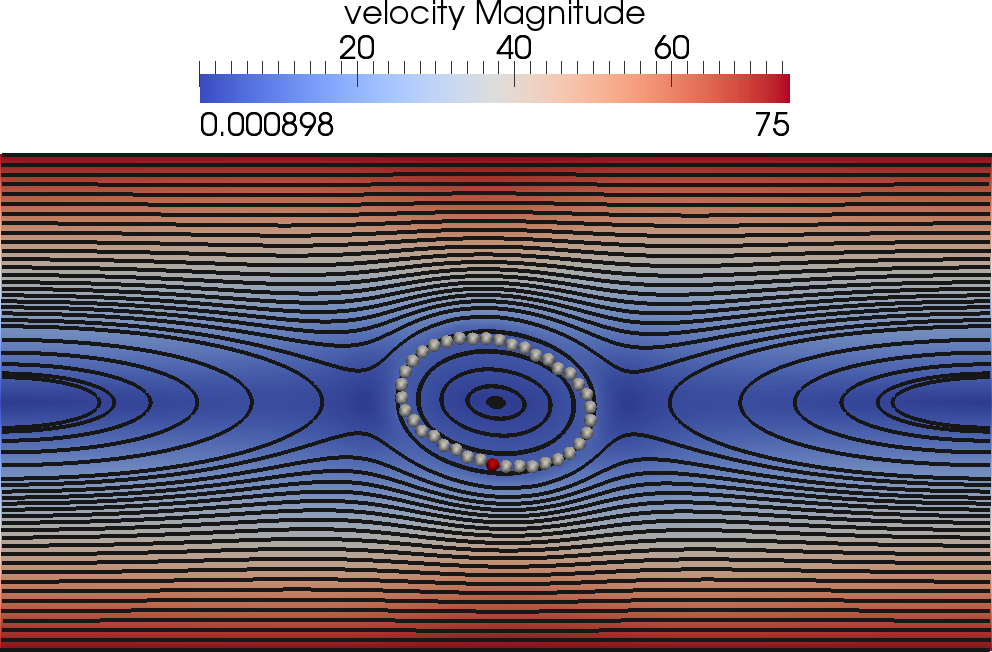}}
  \subfigure[$t=10.5$]{\includegraphics[scale=0.15]{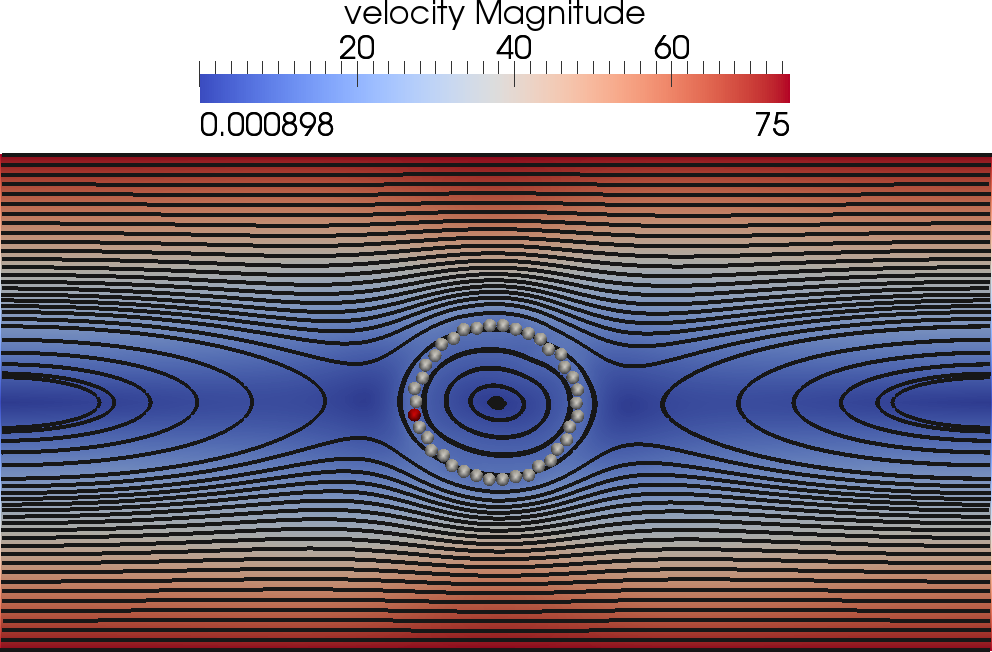}}
  \subfigure[$t=13$]{\includegraphics[scale=0.15]{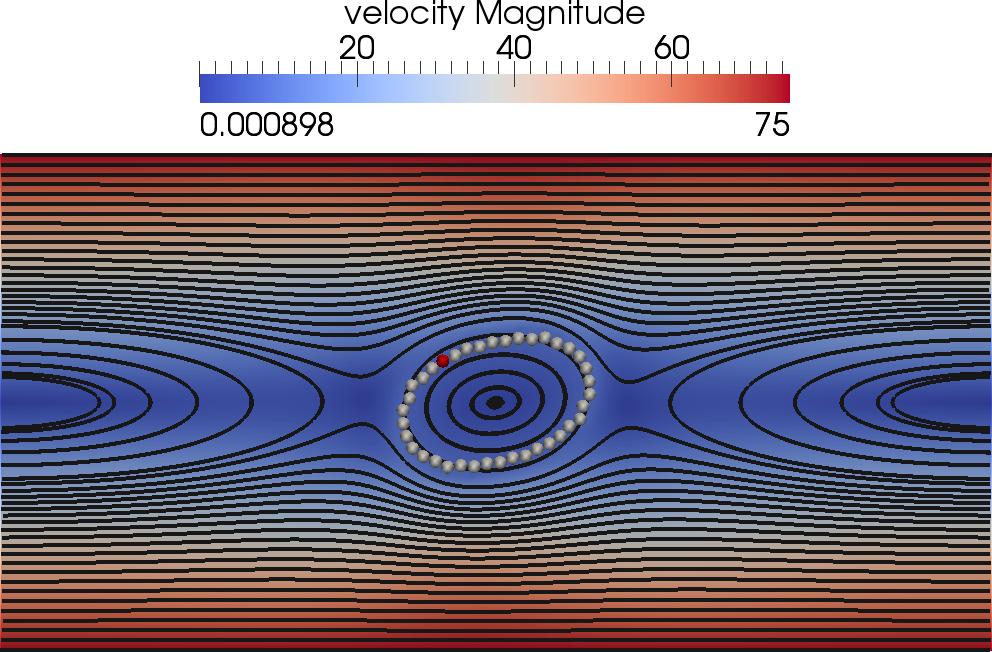}}
  \subfigure[$t=19.5$]{\includegraphics[scale=0.15]{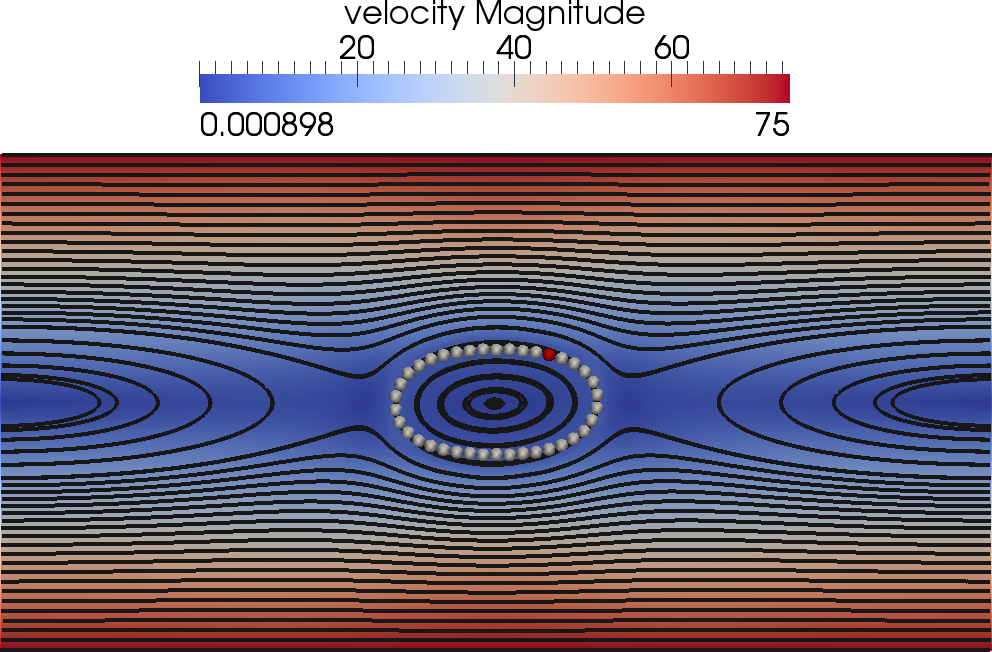}}
  \caption{Streamlines of the velocity field and vesicle position in
    Vacillating-Breathing regime at different time. $\alpha=0.85$,
    $\tau=0.32$, $N=38$ and $\lambda=7.5$.}
  \label{fig:streamVB75}
\end{figure}

Figure \ref{fig:streamVB75} shows the streamlines of the
velocity field, its magnitude and the position of the vesicle at
different times. The initial time corresponds to the first time step
iteration and the final one corresponds to the time during which the
vesicle made half a turn.
In this simulation we take $\lambda=7.5$ as  viscosity ratio and we
keep unchanged all other parameters as in the cases of TT and TB
simulations (the capillary number is then $C_a\simeq 7.68$).

\begin{figure}[ht]
  \centering
   \begin{tikzpicture}
  \begin{axis}[xlabel=Viscosity contrast,ylabel=Angle (degree),
    xmode=normal,
    grid=both, ,minor tick num=4
    ]    
    \addplot table[x=lambda,y=angle]{anglesTTVB08.txt}; 
    \addlegendentry{$\tau=0.2$ $\alpha=0.8$}

    \addplot table[x=lambda,y=angle]{anglesTTVB085.txt}; 
    \addlegendentry{$\tau=0.2$ $\alpha=0.85$}
  \end{axis}
\end{tikzpicture}
  \caption{TT Angles versus viscosity ration.}
  \label{fig:visContrast}
\end{figure}
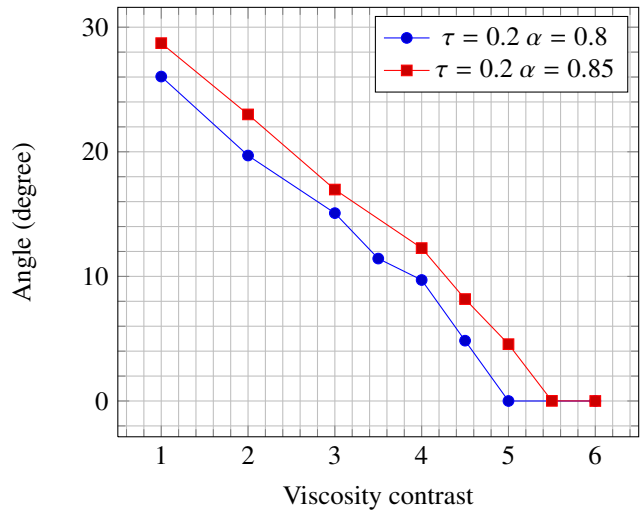

In order to point out the fact that VB is an intermediate regime
between TT and TB, we present in figure \ref{fig:visContrast} the
vesicle angle versus time for different viscosity ratios. We see
clearly that in this case, we have TT regime up to the critical value
$\lambda=5.5$ for which the steady angle is close to zero. Then, for
$\lambda=6.5$ we have a VB motion. Finally TB regime starts from $\lambda=8$.
\begin{figure}
  \centering
  \includegraphics[scale=0.6]{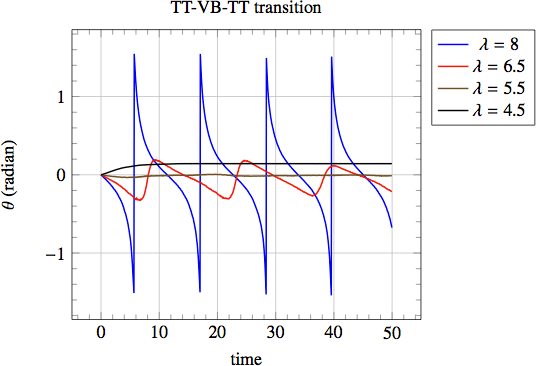}
  \caption{Angles versus time for different viscosity
    ratios. $\alpha=0.85$, $\tau=0.2$.}
  \label{fig:transitionTT-VB-TB}
\end{figure}

For the same purpose, we plot in figure \ref{fig:visContrast} the
steady TT angle versus the viscosity ratio for two values of the
reduced area. One can see that the presence of the VB regime
suppresses the square root singularity at the tumbling threshold
as it is explained in \cite{ghigliotti2010rheology}

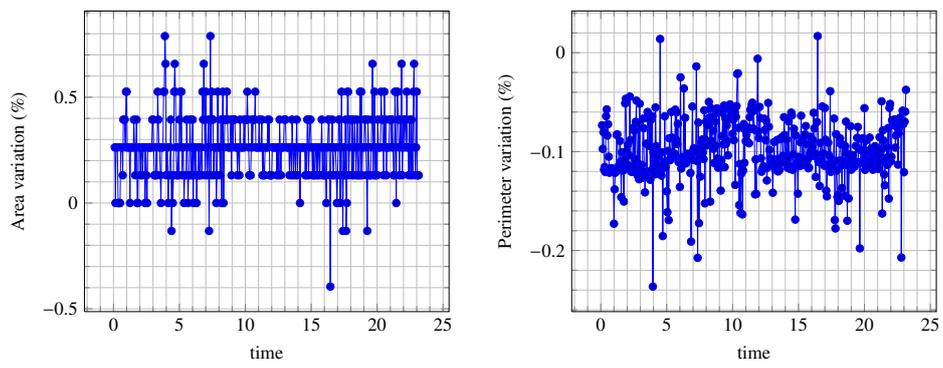
\begin{figure}[hbtp]
  \centering
  \begin{tikzpicture}[scale=0.7]
    \begin{axis}[xlabel=time,ylabel=Area variation (\%), 
      grid=both, ,minor tick num=4,
      ]

      \addplot
      table[x=t,y=V]{VolPerim_VB085.vis07.50.conf0.2.eps5e3-INIT10.txt};

    \end{axis}
  \end{tikzpicture}  \quad
\begin{tikzpicture}[scale=0.7]
  \begin{axis}[xlabel=time,ylabel=Perimeter variation (\%), 
    grid=both, ,minor tick num=4,
    ]

    \addplot
    table[x=t,y=P]{VolPerim_VB085.vis07.50.conf0.2.eps5e3-INIT10.txt};

  \end{axis}
\end{tikzpicture}  
  \caption{Vacillating-Breathing. Area and perimeter versus $t$ for
  $\alpha=0.85$.}
  \label{fig:VB_VolPerim_vs_t-INIT10}
\end{figure}

Finally, we plot in figure \ref{fig:VB_VolPerim_vs_t-INIT10} the area
and perimeter variations in the case of VB regime. Note that we have
the same results in the cases of TT and TB regimes but we choose to
show only the VB case since it is the less favorable regime because the
vesicle undergoes large deformations.

\section{Conclusion}
\label{sec:conclusion}

We introduced in this work a new model to study the dynamics of 2D
vesicles under shear.

First, we described our model and pointed out the way to solve it
using finite elements coupled with a penalty method. Mainly, this
model is based on Stokes equations for the inner and outer fluids and
on a necklace of rigid particles for the vesicle membrane. The
physical properties of the vesicle are taken into account by a couple
of forces acting on the rigid particles to model the inextensibility
and the bending of the membrane. The area and perimeter conservation
are obtained with a good precision by projecting the velocity of the
particles onto a set of admissible fields satisfying these
constraints. These projections are done using Uzawa algorithm.

One of the advantages of our model is its link with fluid/rigid
particles problems. One can see that it is slightly easy to extend any
fluid/rigid particles solver to deal with several vesicles. However, it is not
easy to extend it to 3D vesicles unless to authorize the overlapping
of the rigid particles which introduces technical difficulties.

Second, we have validated our model by retrieving the equilibrium
shapes of vesicles in a fluid at rest and by recovering the TT and TB
regimes.

Finally, we pointed out the transitional regime (when increasing the
viscosity ratio) which is called Vacillating-Breathing or
Swinging. This particular motion has been predicted theoretically in
\cite{PhysRevLett.96.028104} and observed experimentally in
\cite{PhysRevLett.98.128103}. It has been carried out numerically for
3D vesicles in \cite{PhysRevE.83.031921} and for 2D vesicles in
\cite{PhysRevE.80.011901} by adding thermal fluctuations to the
membrane model. To the best of our knowledge, our results are the
first ones recovering numerically the VB regime in 2D without adding
any special ingredient.

\section*{Acknowledgments}
\label{sec:acknowlodgement}

The authors would like to thank B. Maury and C. Misbah for fruitful
discussions and the French National Research Agency (the MOSICOB
project) for the financial support.

\appendix

\section{Numerical bending modulus and capillary number}
\label{sec:appendixCaN}

In this section we derive the numerical capillary number $C_a^N$ which
corresponds to a vesicle's membrane represented by $N$ rigid
particles. 
Let us recall the bending energy $E_b$ we used in our model 
\begin{equation}
  \label{eq:12}
  E_b=\sum_{i}k_{a} (e_i \cdot e_{i+1} + 1).
\end{equation}
We assume that the vesicle has a circular shape with radius $R$. For
large $N$, namely for small $\theta_i^N=\frac{2\pi}{N}$, see figure
\ref{fig:notations_capillaire} for notations, we have
$$
\begin{array}{rcl}
  E_b^N&=&\displaystyle k_a\sum_i (\cos(2\beta_i^N)+1)=k_a\sum_i (\cos(\pi-\theta_i^N)+1),\\
  &=&\displaystyle -k_a\sum_i (\cos(\theta_i^N)-1).\\
\end{array}
$$

Using the fact that $\theta_i^N\approx \frac{2r^N}{R}$, where $r^N=\frac{\pi R}{N}$ is the radius of the particles, one gets for large $N$ :
$$
\begin{array}{rcl}
  E_b^N &\displaystyle\approx&\displaystyle -k_a\sum_i
  \left(-\frac{\left(\theta_i^N\right)^2}{2} \right)\approx k_a\sum_i
  \left(\frac{1}{2} \frac{\left(2r^N\right)^2}{R^2}\right),\\ 
  &\approx&\displaystyle\frac{k_a\,2r^N}{2}\sum_i 2r^N \frac{1}{R^2}\approx \frac{k_a\,2r^N}{2}\int C^2.
\end{array}
$$

From this, together with the fact that the continuous bending energy is given by
$$
E_b=\frac{B}{2}\int C^2,
$$
where $B$ is the bending modulus, we define the discrete bending modulus
$$
B^N=k_a 2r^N=\frac{k_a\,2\pi R}{N}.
$$
The continuous capillary number is defined by (see \cite{Tsubota2010}) :
$$
C_a=\frac{\mu_{out} l^2 R \dot\gamma}{B},
$$
where $\mu_{out}$ is the viscosity of the outer fluid and $l$ is the caracteristic length of
the vesicle. If one takes $l=R$ we obtain
$$
C_a=\frac{\mu_{out} R^3 \dot\gamma}{B}.
$$
We can then define a discrete capillary number : 
$$
C_a^N=\frac{\mu_{out} R^3 \dot\gamma}{B^N}=\frac{\mu_{out} R^3 \dot\gamma}{k_a\,2r^N}.
$$

Note that if we let $k_a$ depend on $N$ and if we chose 
$$
k_a=k_a^N=\frac{B}{2r^N},
$$
we obtain, for all $N$
$$
B^N=B\quad \hbox{and}\quad C_a^N=C_a.
$$

\bibliographystyle{elsarticle-num-names}
\bibliography{necklace}

\end{document}